\documentclass[reprint,aps,prb,amsmath,amssymb]{revtex4}

\usepackage{graphicx}
\usepackage{exscale}
\usepackage{subfigure}

\usepackage{amsfonts}
\usepackage{amsmath}
\usepackage{amssymb}
\usepackage{subfigure}

\newcommand{\ket}[1]{\left| #1 \right\rangle}

\newcommand{\be}{\begin{equation}}
\newcommand{\ee}{\end{equation}}
\newcommand{\bea}{\begin{eqnarray}}
\newcommand{\eea}{\end{eqnarray}}
 
\newcommand\scalemath[2]{\scalebox{#1}{\mbox{\ensuremath{\displaystyle #2}}}}

\begin{document}

\title{Exciton propagation   via  quantum walks based on non-Hermitian coin flip operations}

\author{A. Thilagam}
\email {thilaphys@gmail.com}
\affiliation{\rm Information Technology, Engineering and Environment, \\
University of South Australia, \\ South Australia  5095 \\ 
Australia}

\begin{abstract}
We examine  the coherent propagation of the one-dimensional 
Frenkel exciton (correlated electron-hole pair system) 
based on a model of a quantum walker in multi-dimensional Hilbert space. 
The walk is   governed by a  non-Hermitian coin flip operation
 coupled to a generalized shift process. 
The dissipative coin flip operation is associated with  
amplitude leakages at  occupied sites, typical 
of processes which occur when an exciton is
transferred along dimer sites in  photosynthetic protein complexes.
We analyze the  characteristics  probability distribution of
the one-dimensional quantum walk for various system parameters, and
examine  the complex interplay  between non-Markovian signatures  and
amplitude leakages within  the  Hilbert position subspace.
The visibility of  topological defects such as exceptional points, and non-Markovian signatures
 via quantum tomography based spectroscopic measurements is discussed.
\end{abstract}

\maketitle
\section{Introduction}

It is well known that systems which enable quantum walks of one or more
propagating entities (quasi-particle or excitation)
exhibit intriguing features not seen in the classical 
walker \cite{advan,advan1,advan2,advan3,advan3b,shi}.
For instance, the  quantum walker  moves away faster from the point of initial start 
than the classical random walker.
Several works \cite{viv,under,child} have shown that 
the quantum information encoded in the walker's positions provide a basis for
universal quantum computation, even allowing single-qubit measurements to be incorporated 
with localized operators and existing quantum error-correcting codes
\cite{under}. Quantum walks can be  modeled either as  continuous-time quantum walk
(CTQW)  \cite{farhi} or as discrete-time quantum walk (DTQW) \cite{advan1,meyer,shi2}.
While  the  system evolves with no restriction on time during CTQW,
the DTQW involves an entangled system of a walker and a coin, and 
 an evolution operator which is applied to both systems in discrete time steps.
In both the CTQW and DTQW models, the quantum walker moves along discrete points of
the  undirected graphs.

Since the inception of the basic  model of a sole walker moving 
left or right simultaneously, and conditional on the  two outcomes of a flipped quantum coin,
several variants of the  DTQW have been proposed. One known model is  based on 
  multiple walkers and coins, \cite{prop1,prop2,prop3}
the second, centers on the presence of a distinguished node due to local change 
of phase, \cite{node} while  a third DTQW
variant involves quantum  walks in the presence of traps \cite{trap}. 
A non-Hermitian coin toss model has also been  considered on lattices  and finite graphs,
incorporating the behavior of open quantum systems \cite{attal}. 
Quantum walks can be categorized depending on the behavior of the spreading measure $\sigma$ 
in the long time limit, accordingly the walk can be  ballistic ($\sigma \propto t$), diffusive ($\sigma \propto \sqrt{t}$)
or localized ($\sigma \propto$ const) \cite{cate0,cate,cate2}.
Quantum walks  have been demonstrated using single photons in space, with
walks  up to six steps measured, and the quantum-to-classical transition  analyzed via the
decoherence mechanism \cite{expt1}. Another study \cite{expt2} examined
 the  simulations of quantum walks via confinement of a single $Mg^+$ ion in a linear Paul trap.
Recently \cite{expt3} quantum walks were performed 
on an integrated array of symmetric, polarization insensitive, directional 
couplers in which two-photon polarization entangled states were introduced. 
The dynamics of symmetric and antisymmetric particles undergoing quantum walks
was demonstrated experimentally by utilizing the different statistics of singlet and triplet
entangled states  \cite{expt3}.

The various  modifications and categorization of quantum walks provide
useful techniques in simulating realistic systems such as
photosynthetic protein complexes \cite{engel,adolp,flem,olb,caru,photo,ishi} which convert solar energy
 into chemical energy through extremely efficient propagation schemes. 
The energy of photons absorbed by a intricate network of 
bacteriochlorophylls is transferred,  with almost near unity quantum
efficiency to a reaction center (RC) where energy conversion  and 
charge separation occurs. The well-known Fenna$-$Mathews$-$Olson (FMO) pigment protein complex 
acts an interface between the chlorosomic antenna and the 
reaction center in the green sulphur bacteria \cite{adolp,flem}. The  FMO interface constitutes  eight 
coupled bacteriochlorophyll-a chromophores (BChl a) which gives rise to
eight excitonic states that perform the  vital role of transferring
energy from the antennae end toward the reaction center.

Various theories \cite{caru,ishi,silbey,thila1,thila2,pleniorev,thilarev,reben2}  have been proposed in 
recent years to account for the high
efficiencies of energy transfer in the FMO protein complex. 
Ishizaki  et. al. \cite{ishi} showed that quantum coherence allows excitations to propagate without being
trapped at sites with  lower energies, while  the role of a
critical ratio in  quantum coherence and environmental noise 
in achieving  optimal  functionality and  efficiency
 in photosynthetic  systems was examined in other  works  \cite{caru,reben2}. 
In an earlier work, we demonstrated that quantum information processing tasks involving teleportation 
and decodification based on specific states ($W$-like states) of the FMO complex may  contribute to
 coherent oscillations at physiological temperatures \cite{thila2}.
Recently, we considered the importance of the Zeno mechanism-non-Markovianity link  \cite{thila3}
in  wider networked molecular systems of
 photosynthetic membrane surfaces that constitute thousands of bacteriochorophylls.
While  protein-induced (time-dependent) fluctuations and dissipation  play an important role in controlling the energy transfer dynamics
in photosynthetic pigment-protein complexes, propagation schemes based on quantum coherence
has gained attention in recent years \cite{caru,pleniorev,thilarev,reben2}.
The observations of long-lasting coherence at ambient temperatures \cite{engel} 
remains to be fully accounted for, either by semi-classical or quantum approaches,
or a combination of both. Accordingly, 
the exact details of  the  link between the robust coherence oscillations and the ultrafast 
times taken for  excitation energy to be
transferred  from the antenna to the reaction center
in the Fenna-Matthews-Olson trimer  remains largely unidentified.

One approach that remains to be fully exploited in relation to the  examination of  the long-lasting coherences
in light-harvesting systems,  involves the non-equilibrium dynamics
 of  non-Hermitian  components in open quantum dissipative systems.
The presence of  non-Hermitian attributes  gives rise to the characteristic features of quantum   states 
 with finite lifetimes, and quasi-bound state resonances that are  present in the
continuum partition.  In the conventional Hermitian case,
states remain un-coupled to the continuum states, and possess  infinitely long lifetimes.
The real and imaginary components  of the non-Hermitian 
 eigenfunctions evolve almost independently of each other during 
 avoided level crossing \cite{rotter}. These components also give rise to 
 dynamical phase transitions that involves
a bifurcation of the time scales corresponding to the lifetimes of the resonance states.
The phase transitions arising from  changes in environmental parameters
are linked to topological defects known as exceptional points.
The salient feature of short-lived and  long-lived quantum states arising from
non-Hermitian dynamics has important consequences for the
 non-ideal exciton arising from  the Pauli exclusion between  electrons and holes \cite{comb}.
Pauli exclusion processes operate independently of the widely known coulomb processes 
 in interacting fermion systems, and  contribute to hermiticity
of excitonic Hamiltonians. The non-ideal bosonic features of excitons was examined using
a non-Hermitian open quantum system approach recently \cite{thilmath},
where it was shown that long-lived quantum coherence in photosynthetic
 complexes may be assisted by small bosonic deviation measures.
Moreover the involvement of a higher number of excitons  during energy exchanges
could help realize  a highly  correlated molecular environment.

In this work, we examine the characteristics of 
a one-dimensional quantum walk where there is possibility
of loss of amplitude at newly located site due to a non-Hermitian coin flip operation.
The leakages in excitonic occupation probabilities can be attributed to a 
combination  of  factors such as those arising from
dissipation  due to exciton-phonon interactions, exciton-exciton annihilation, presence of trapping centers
and  the ubiquitous presence of Pauli interactions  as a result of the  composite nature of excitonic systems \cite{pauli}.
In connection with the quantum walk on molecular sites, we focus on 
  two signatures of exciton dynamics that can be derived from 
 a  quantum information theoretic approach to spectroscopic experiments.
The first  relates to the appearance of topological defects (exceptional points) due to 
coalescing of operator eigenvalues at specific values of the
system parameters. The second feature relates to  
non-Markovian dynamics  that arise at short time-scales or stronger system-environment couplings,
resulting in  violation of complete positivity of the density matrix of the probed
quantum system. The detection of these features have  importance in the relatively new field of 
quantum process tomography (QPT) \cite{wein,walm,milota,excitomo}
where the  quantum state amplitudes in the Hilbert space
are obtained via a time-series  of frequency-frequency
correlation spectroscopic results. 

This  paper is organized as follows. In Section \ref{basics},
we  analyze  the quantum walk on a line with a non-dissipative coin operator
coupled to a generalized shift operation, and examine
the  probability distribution profile  with respect to system
parameters. In Section \ref{nonhermi}, we extend the problem to include
the quantum walk based on a non-Hermitian coin flip operation which is used
to examine the propagation of exciton in a simple one-dimensional system 
in Section \ref{excitrans}. This
model has relevance in light-harvesting systems as well as other bio-molecular networks such as J-aggregates.
 The characteristic features  of the probability distribution
with respect to changes in the waiting time $\tau$ (either walk or wait time) and 
$\lambda$ (dissipative measure) is analyzed with respect to the 
 total number of steps taken by the quantum walker. A discussion of the implementation
of the non-Hermitian coin flip operation via the exciton spin state is 
provided in Section \ref{flip} and the influence of the dissipative measure $\lambda$ 
on the Von Neumann entropy  is analyzed in Section \ref{von}.
In the final Section  \ref{meaQW}, we briefly review recent studies
of  quantum process tomography of relevance to excitonic systems,
and present results associated with  
  two signatures of exciton dynamics: (1) exceptional points, and (b) non-Markovianity.
Section \ref{con}  provides our summary and some conclusions.

\section{Basics of the discrete-time quantum walk}\label{basics}

The one-dimensional discrete time quantum walk takes place
in the total Hilbert space, ${\cal H}_w$=${\cal H}_p\otimes{\cal H}_c$
where ${\cal H}_p$ (${\cal H}_c$) denotes the position (coin) Hilbert space.
 ${\cal H}_p$ is spanned by the basis states $|m_p\rangle$ of the walker, 
${\cal H}_P$=$\textrm{Span}\left\{|m\rangle|\ m \in \textsl{Z}\right \}$ while 
${\cal H}_c$ is spanned by the coin states $|1_c\rangle,|0_c\rangle$.
A conditional shift operator is employed  such that the walker shifts  
one step forward if the linked coin state occupies a specific basis state
 (e.g. $|0_c\rangle$), or one step backwards if 
the coin state occupies the alternative  basis state (e.g. $|1_c\rangle$). 
The operator which acts on the total Hilbert space, ${\cal H}_w$ for
a quantum walk involving  a  single step  is given by the propagator,
\begin{equation}
{\hat U} = {\hat S}  \cdot \left(\mathbb{{\hat I}}_p \otimes {\hat C}\right).
\label{qwt}
\end{equation}
where $\mathbb{{\hat I}}_p$ is 
the unit operator that acts on  $\mathcal{H}_P$.  A typical  displacement operator ${\hat S}$ has the form
\begin{equation}
\label{shift_one}
{\hat S} = \sum\limits_{m=-\infty}^{+\infty}|(m+1)_p\rangle \langle m_p|
\otimes|0_c\rangle \langle 0_c|+\sum\limits_{m=-\infty}^{+\infty}|(m-1)_p\rangle \langle m_p|\otimes|1_c\rangle \langle 1_c|.
\end{equation}

Prior to the application of the displacement ${\hat S}$,
a coin flip operator ${\hat C}$ that acts on $\mathcal{H}_C$ is applied.
${\hat C}$ may be an arbitrary unitary operator of   the form
\begin{equation}
{\hat C}(\alpha) = \left [
\begin{array}{cc}
{\alpha} & \sqrt{1-\alpha^2} \\
\sqrt{1-\alpha^2} & -{\alpha} \\
\end{array}
\right].
\label{cos}
\end{equation}
where $\alpha=\frac{1}{\sqrt 2}$ is associated with the well known  Hadamard walk.
The discrete quantum walk after $t$ steps  is given by
\begin{equation}
|\Psi \rangle_t = ({\hat U})^t  |0_p \rangle \otimes |0_c \rangle,
\label{succint_quantum_walk}
\end{equation}
where $|0_p \rangle$($|0_c \rangle$) denote the initial
state of the quantum walker (coin).

\begin{figure}[htp]
  \begin{center}
\subfigure{\label{aa}\includegraphics[width=7cm]{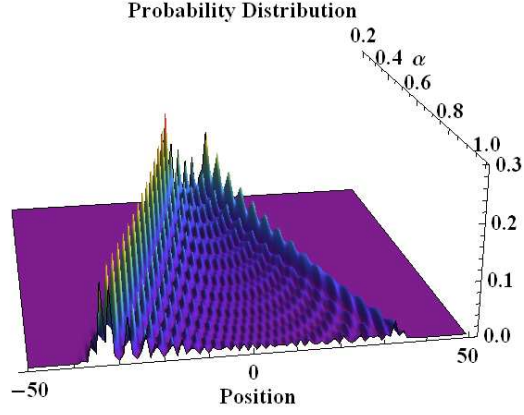}}\vspace{-1.1mm} \hspace{1.1mm}
     \end{center}
\caption{Probability distribution of a quantum walk as function of position and 
$\alpha$  obtained after 50 steps, and based on the coin operator in Eq.(\ref{cos}).
}
\label{qwalkg}
\end{figure}

In this work, we consider a generalized shift operation where
\bea
\label{gshift}
\ket{0_c,m_p} &\rightarrow& \frac{1}{\sqrt 2} \ket{0_c,(m-1)_p}+\frac{1}{\sqrt 2} \ket{1_c,m_p}, \\
\ket{1_c,m_p} &\rightarrow& -\frac{1}{\sqrt 2} \ket{0_c,m_p}+\frac{1}{\sqrt 2} \ket{1_c,(m+1)_p}.
\eea
The action of ${\hat C}(\alpha)$ (Eq.(\ref{cos})) in the propagator ${\hat U}$ (Eq.(\ref{qwt}))
will produce a  first step configuration from an initial state, $\ket{0_c,0_p}$, as follows
\bea
\ket{0_c,0_p} \longrightarrow -\frac{\sqrt{1-\alpha^2}}{\sqrt{2}} \left|0_{c},0_{p}\right\rangle
   +\frac{\sqrt{1-\alpha^2}}{\sqrt{2}}
   \left|1_{c},1_{p}\right\rangle+\frac{\alpha }{\sqrt{2}}
   \left|0_{c},-1_{p}\right\rangle+\frac{\alpha}{\sqrt{2}}
   \left|1_{c},0_{p}\right\rangle 
\label{one}
\eea
After two steps, the initial configuration changes to
\bea
\label{two}
\ket{0_c,0_p} \longrightarrow &&
\frac{1}{2} \alpha^2 \left|0_{c},-2_{p}\right\rangle +(1-\frac{1}{2}
   \alpha^2) \left|0_{c},0_{p}\right\rangle +\frac{1}{2} \alpha^2
   \left|1_{c},-1_{p}\right\rangle  \\ &-& \nonumber \frac{1}{2} \alpha^2
   \left|1_{c},1_{p}\right\rangle  -\frac{1}{2} \alpha \sqrt{1-\alpha^2}
    \left|0_{c},-1_{p}\right\rangle 
+\frac{1}{2}  \alpha
   \sqrt{1-\alpha^2} \left|0_{c},1_{p}\right\rangle \\ \nonumber
   &+& \frac{1}{2}  \alpha \sqrt{1-\alpha^2}
   \left|1_{c},0_{p}\right\rangle  -\frac{1}{2}\alpha \sqrt{1-\alpha^2}
    \left|1_{c},2_{p}\right\rangle
   \eea
The  expression for the distribution profile of the quantum walk becomes longer  with 
increasing number of steps. For the one- and two- step
cases (Eqs.(\ref{one}),(\ref{two})), it  can be easily verified that  the sum of probabilities 
of occupation at all sites equals unity. In general this rule applies to  any number of
steps taken by the walker, provided there is absence of  dissipation at the occupied
sites. In  Fig.~\ref{qwalkg}, we note the spreading out of the probability distribution,
typified by two counter-propagating dissimilar peaks,
with increase in $\alpha$  which controls the coin operator in Eq.(\ref{cos}).
The probability distribution peaks increase in height as $\alpha$ decreases. 
Due to the constructive and destructive
interference features of  a quantum walk,
the position coordinate spreads out faster  in time compared to the classical random walk,
as is well known.
A symmetrical profile in the probability distribution
can be obtained by changing the
initial state, $\ket{0_c,0_p}$ to initial state, $\frac{1}{\sqrt 2}(\ket{0_c,0_p}+ i \ket{1_c,0_p})$.
In the next Section, we consider a non-Hermitian walk where there is 
loss of amplitude at a newly occupied site.

\section{Quantum walk based on a non-Hermitian coin flip operation}\label{nonhermi}

We consider the non-Hermitian  coin flip operator ${\hat C'}$ of the following form
\begin{equation}
{\hat C'}(\alpha) = \left [
\begin{array}{cc}
{\alpha_1} & \alpha_2 \\
\alpha_2 & -{\alpha_1} \\
\end{array}
\right].
\label{cosH}
\end{equation}
where $\alpha_2 < \alpha_1$ and  $\alpha_1^2$+$\alpha_2^2 < 1$, with the difference
$\Delta$ = $\alpha_1-\alpha_2$ determined by parameters specific to the system undergoing
non-Hermitian dynamics.
The action of ${\hat C'}$ in the propagator ${\hat U}$ (Eq.(\ref{qwt}))
will produce a  first step configuration from an initial state, $\ket{0_c,0_p}$, as follows
\bea
\ket{0_c,0_p} \longrightarrow \frac{\alpha _1 \left|0_{c},-1_{p}\right\rangle }{\sqrt{2}}-\frac{\alpha _2
   \left|0_{c},0_{p}\right\rangle }{\sqrt{2}}+\frac{\alpha _1
   \left|1_{c},0_{p}\right\rangle }{\sqrt{2}}+\frac{\alpha _2
   \left|1_{c},1_{p}\right\rangle }{\sqrt{2}}
\label{onen}
\eea
where the sum of probabilities 
of occupation at all  possible sites after one step, $N_1=(\alpha _1^2+ \alpha _2^2)$.
After two steps, the initial configuration changes to
\bea
\label{twon}
\ket{0_c,0_p} \longrightarrow &&
\frac{1}{2} \alpha _1^2 \left|0_{c},-2_{p}\right\rangle + (\frac{1}{2} \alpha _1^2+\alpha _2^2)
   \left|0_{c},0_{p}\right\rangle +\frac{1}{2} \alpha _1^2
   \left|1_{c},-1_{p}\right\rangle -\frac{1}{2} \alpha _1^2
   \left|1_{c},1_{p}\right\rangle  \\ &-& \nonumber
\frac{1}{2} \alpha _2 \alpha _1
   \left|0_{c},-1_{p}\right\rangle +\frac{1}{2} \alpha _2 \alpha _1
   \left|0_{c},1_{p}\right\rangle +\frac{1}{2} \alpha _2 \alpha _1
   \left|1_{c},0_{p}\right\rangle - \frac{1}{2} \alpha _2 \alpha _1
   \left|1_{c},2_{p}\right\rangle 
   \eea
where the sum of probabilities 
of occupation at all possible sites after two steps, $N_2=(\alpha _1^2+ \alpha _2^2)^2$.
The expression for $N_m$ for  the general $m$ steps decreases
monotonically with increase in $m$ for $\alpha_2 < \alpha_1$ as
$N_m=(\alpha _1^2+ \alpha _2^2)^m$. 

The form of the multipartite state in the total Hilbert space,
${\cal H}_p\otimes{\cal H}_c$, is dependent on 
the coefficients $\alpha _1,\alpha _2$ and the number of steps taken by the
walker. We briefly examine the density matrix in the coin Hilbert space
after three steps ($N=3$) as the higher step cases involve longer expressions.
By tracing  out the position states of the walker, we obtain
a   reduced density matrix describing the quantum walk
in the basis $(\ket{0_{c}},\ket{1_{c}})$
\be
\label{matc}
\rho_c(\alpha_1,\alpha_2) =
\left(
\begin{array}{cc}
 \frac{1}{4} \left(\alpha_1^2+\alpha_2^2\right) \left(3 \alpha_1^4+8 \alpha_2^2 \alpha_1^2+2 \alpha_2^4\right) & \frac{1}{4} \alpha_1 \alpha_2^3 \left(\alpha_1^2-4 \alpha_2^2\right) \\
 \frac{1}{4} \alpha_1 \alpha_2^3 \left(\alpha_1^2-4 \alpha_2^2\right) & \frac{1}{4} \left(\alpha_1^2+\alpha_2^2\right) \left(\alpha_1^4+2 \alpha_2^4\right) \\
\end{array}
\right)
\ee
For the simple case of $\alpha _1 = \alpha _2 = \frac{1}{\sqrt{2}}$,
we get unit trace with non-negative eigenvalues, $(0.82626, 0.17374)$.
For the case of $\alpha _1 =r;  \alpha _2 = \sqrt{1-r^2}$,
we again obtain non-negative eigenvalues, $e_1=
\frac{1}{4}(2- f(r), e_2= \frac{1}{4} (2 + f(r))$
where $f(r)=[r^2 \left(-25 r^{10}+115 r^8-202 r^6+169 r^4-72 r^2+16\right)]^{1/2}$,
thus $e_1+e_2$=1. However for the dissipative scenario where $\alpha _1=r;  \alpha _2 = \sqrt{1-r^2-t}$,
we obtain $e_1+e_2$=$\left(1- t^3+3 t^2-3 t\right)$ which 
 decreases monotonically with increase in $t$. The sum of the eigenvalues 
of reduced density operator may used as a rough indicator of whether the trajectories of the quantum 
walker has indeed a physically realistic interpretation.

\section{Exciton transport and the dissipative non-Hermitian coin flip operation}\label{excitrans}
The  propagation of the exciton (electron-hole pair system) is generally not
one-dimensional in light-harvesting and other bio-molecular systems such as J-aggregates,
as evidenced in the volume of space transversed by the excitation.
In the coherent transport model, the exciton moves as a superposition
of  delocalized excitations over the  crystal space as an extended entangled system,
and accounts well for the large free exciton path during coherent migration. In this case,
the exciton propagation mechanism  may be viewed as similar to that of  a quantum walker
on a graph  where each vertex point represents a lattice site \cite{thilapra},
with propagation route determined by the edge joining one vertex to another,
Obviously, the number of edges incident on a specific vertex is
considerably simplified for one-dimensional systems. It is possible
to single out a particular vertex as playing the role of a trapping center
where energy is collected as is the case in the 
reaction center (RC) of the green sulphur bacteria \cite{adolp,flem},
and in which  energy conversion occurs.

In order to examine the role
of systems parameters such as dissipation rate and time spent at a
visited site, we consider the one-dimensional quantum walker
for a fixed number of sites. We assume a coin flip operation based on the  prototypical 
dissipation process that takes place during transfer of excitation  between two sites  $\ket{l}$
and $\ket{m}$ with a  total Hamiltonian of the form ($\hbar$=1)
\be
\label{pheq}
\hat{H}_{ex}= E_l  B_{ l}^\dagger B_{ l}+ E_m B_{ m}^\dagger B_{ m}+
V B_{ l}^\dagger B_{ m}+ V B_{ m}^\dagger B_{l}
- i\gamma_{l}  B_{ l}^\dagger B_{ l}- i\gamma_{m}   B_{m}^\dagger B_{m}
\ee
where $B_{ l}^\dagger$ ($B_{ m}$) is the creation (annihilation) operator of
energy of magnitude $E_l$ ($E_m$)  at site $l$ ($m$). Eq.(\ref{pheq})
may represent transfer of excited states  between neighboring  sites  in
 photosynthetic protein complexes of light-harvesting systems and
quantum J-aggregates of organic dye molecular systems.
$V$ denotes  the intersite tunneling energy which  is
taken to be real and positive,  and the dissipative terms $\gamma_{l}$ ($\gamma_{m}$) 
denote the leakage rate of energy at site $l$ ($m$). Eq.(\ref{pheq}) can be solved
via the retarded Green's function of the form
\be
G_{l,m}(t)=-i {\rm \Theta(t)} \langle \{B_{ l}(t),B_{ m}^\dagger(0)\}\rangle
\label{geq}
\ee
where $\Theta(t)$ denotes a step function. The Fourier transform, 
$G_{l,m}(E)$=$\int_{-\infty}^{\infty} dt\, G_{l,m}(t)\, e^{iEt}$ 
is given by \cite{lipp}
\be
\label{hermg}
G_{l,m}^{-1}(E)= \left[ \begin{array}{cc} E-E_l+i \eta+ & -V \\ -V & 
E-E_m+i \eta+ \end{array} \right]+ \frac{i}{2} \left( \begin{array}{cc} \gamma_{l} & 0 \\ 0 & 
\gamma_{m} \end{array} \right).
\ee
where $\eta$ is a very  small number. For  an initial state 
in which  the  excitation is present only at site $l$, the
 probability $P_{l,m}$  of excitation transfer from  site $l$ to $m$ is
is obtained by inverting Eq.(\ref{hermg})
\be
P_{l,m}(t)=\frac{2V^2}{|\Omega|^2} e^{-\gamma_d t} 
(\cosh\Omega_i t - \cos\Omega_r t),
\ee
where $\gamma_d$=$\frac{1}{2}(\gamma_{l}$+$\gamma_{m})$, $\Omega \equiv \Omega_r + i\Omega_i\equiv\sqrt{4V^2 + 
(E_0-i \bar{\gamma_d})^2}$, $E_0$=$E_m$-$E_l$,$\bar{\gamma_d}$=
$\frac{1}{2}(\gamma_{m}$-$\gamma_{l})$. The analytical form for the probability,
$P_{l,l}(t)$ is lengthy and is not provided here. Simpler expressions can be obtained for
the probabilities, $P_{l,m}(t)$ and $P_{l,l}(t)$ at  the resonance frequency where $E_0$=0 as $E_m$=$E_l$.
At the  coherent tunneling  regime ($V_0 > \frac{\gamma_d}{2}$) we obtain
\be
\label{cor1}
P_{ll} =  e^{-\gamma_d t}
\left[\cos{\Omega_0} t- \frac{\bar{\gamma_d}}{2 \Omega_0}\sin{\Omega_0} t \right]^2, \quad \quad
 P_{lm} =   e^{-\gamma_d t} \frac{V^2}{\Omega_0^2}\sin^2{\Omega_0} t,
\ee
where   $\Omega_0=\sqrt{4V^2 - \bar{\gamma_d}^2}$.
The total probabilities, $P_{ll}$+$P_{lm} \le 1$ with
the loss of normalization  dependent on the dissipation terms, $\gamma_d$
and $\bar{\gamma_d}$. For the case of 
 incoherent tunneling ($V_0 < \frac{\gamma_d}{2}$), we replace $\sin[x]$ ($\cos[x]$) by   $\sinh[x]$
($\cosh[x]$). A topological defect occurs at the exceptional point,
 $\Omega_0 = 0, \; V = \frac{\bar{\gamma_d}}{2}$,  which is a unique feature of non-Hermitian quantum systems.
At the exceptional point,  two eigenvalues of an operator coalesce
as a result of a change in  system parameters, with  the
 two mutually orthogonal  states merging into one self-orthogonal state.
We next employ the expressions in Eq.(\ref{cor1})
as terms which appear in the non-Hermitian coin flip operation, ${\hat C'}(\alpha)$ (see Eq.(\ref{cosH}))
\bea
\label{sub}
\alpha_1 = e^{-\lambda /4 \tau}
\left[\cos{\Omega_0} \tau- \frac{\lambda}{4 \Omega_0}\sin{\Omega_0} \tau \right],\quad \quad
\alpha_2 =   e^{-\lambda /4 \tau} \frac{V}{\Omega_0}\sin{\Omega_0} \tau
\eea
where $\tau$ can be considered as the time taken to make one quantum walk. We set $\gamma_{m}$ =$\lambda$,
$\gamma_{l}=0$ for simplicity in numerical analysis.

\begin{figure}[htp]
  \begin{center}
\subfigure{\label{aa}\includegraphics[width=6.3cm]{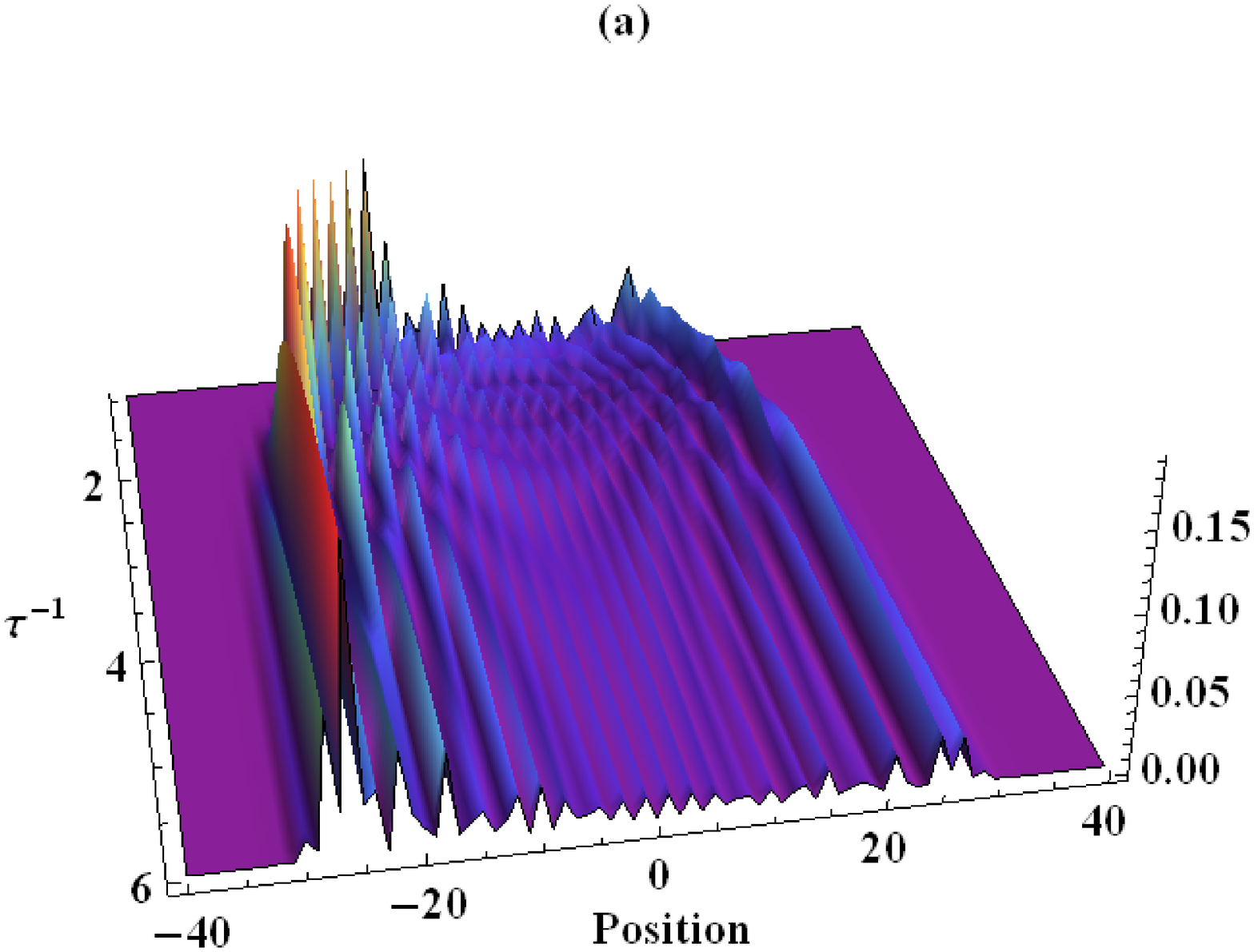}}\vspace{-1.1mm} \hspace{1.1mm}
\subfigure{\label{ab}\includegraphics[width=6cm]{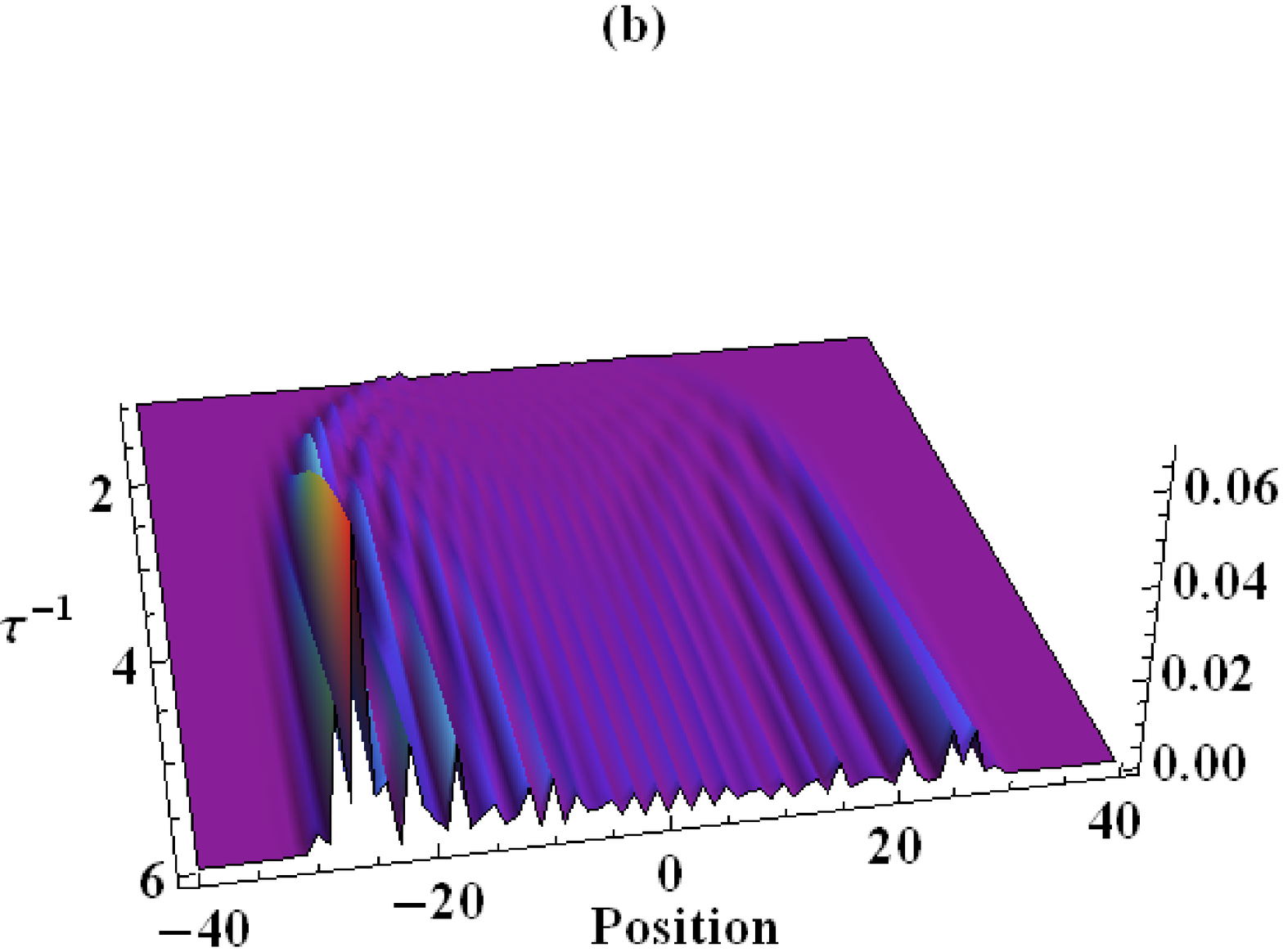}}\vspace{-1.1mm} \hspace{1.1mm}
     \end{center}
\caption{Probability distribution of a quantum walk (z-axis) as function of position and 
$\tau^{-1}$  obtained after 40 steps at (a) $\lambda$ =0, (b) $\lambda$ =0.15
using the coin operator in Eq.(\ref{cosH}) with parameters given in 
Eq.(\ref{sub}). A unit system in which $\hbar$=1,  $\Omega_0$=1 is used.
}
\label{qwalkt}
\end{figure}

\begin{figure}[htp]
  \begin{center}
\subfigure{\label{aa}\includegraphics[width=4.2cm]{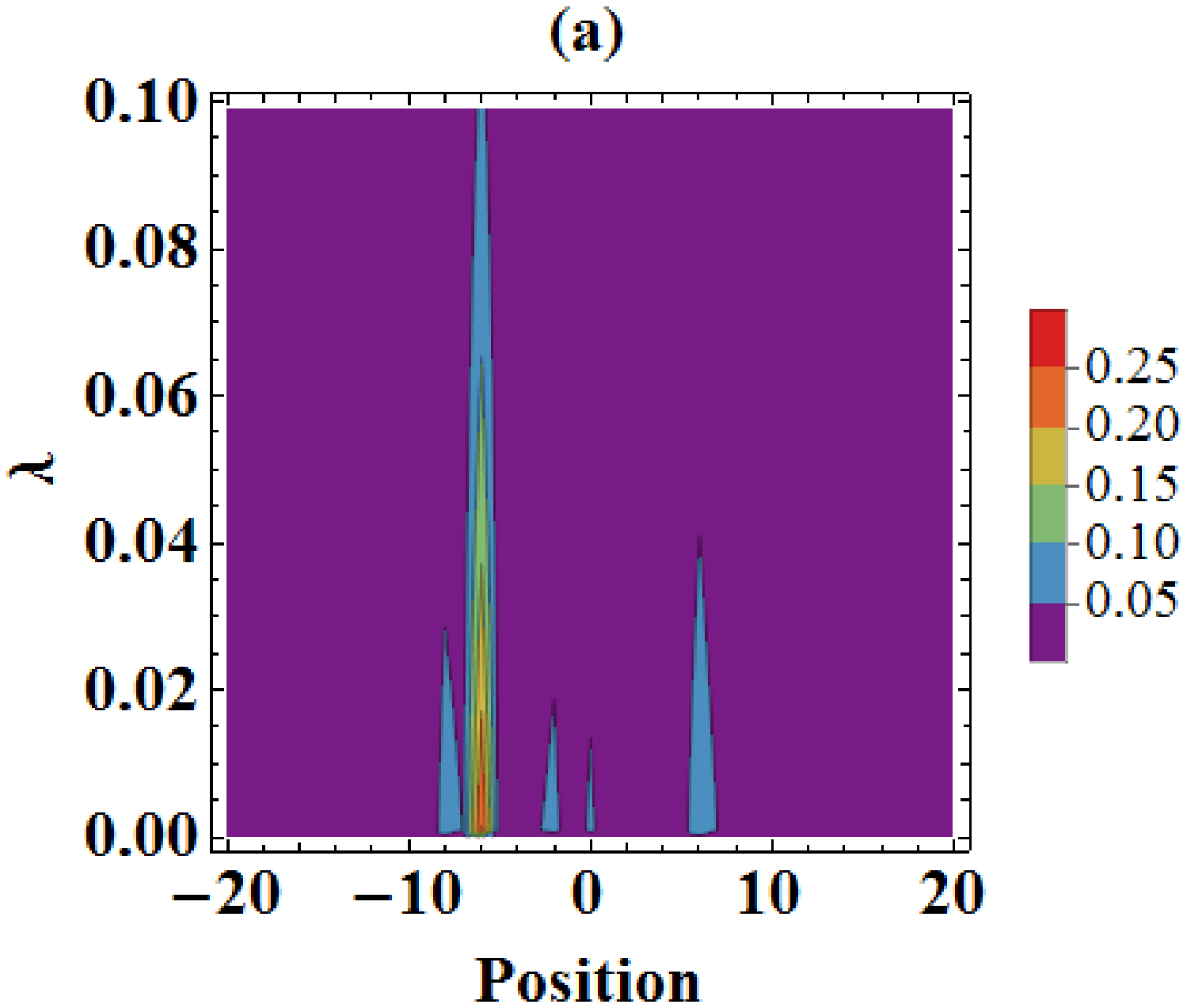}}\vspace{-1.1mm} \hspace{1.1mm}
\subfigure{\label{ab}\includegraphics[width=4.2cm]{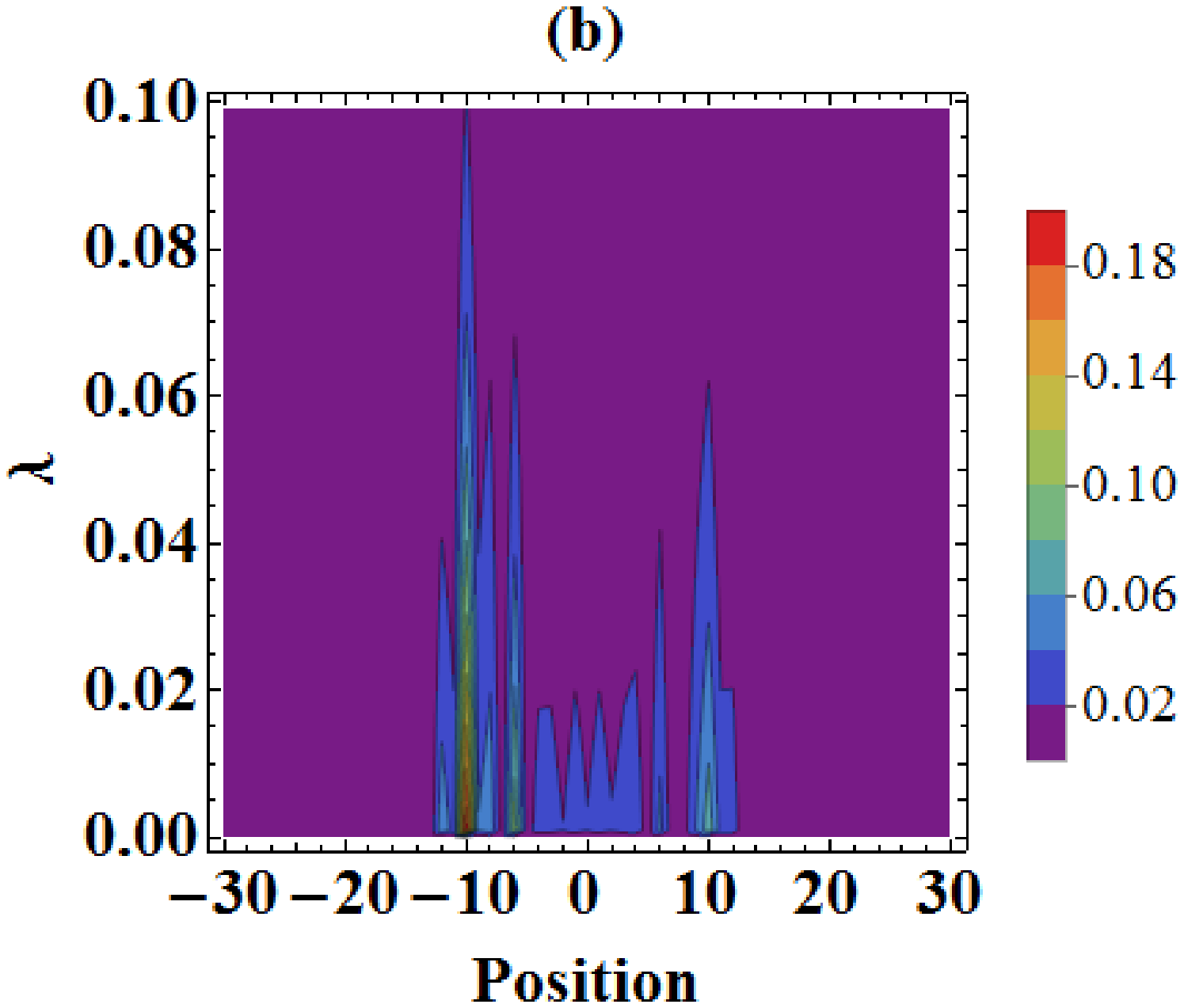}}\vspace{-1.1mm} \hspace{1.1mm}
\subfigure{\label{aa}\includegraphics[width=4.2cm]{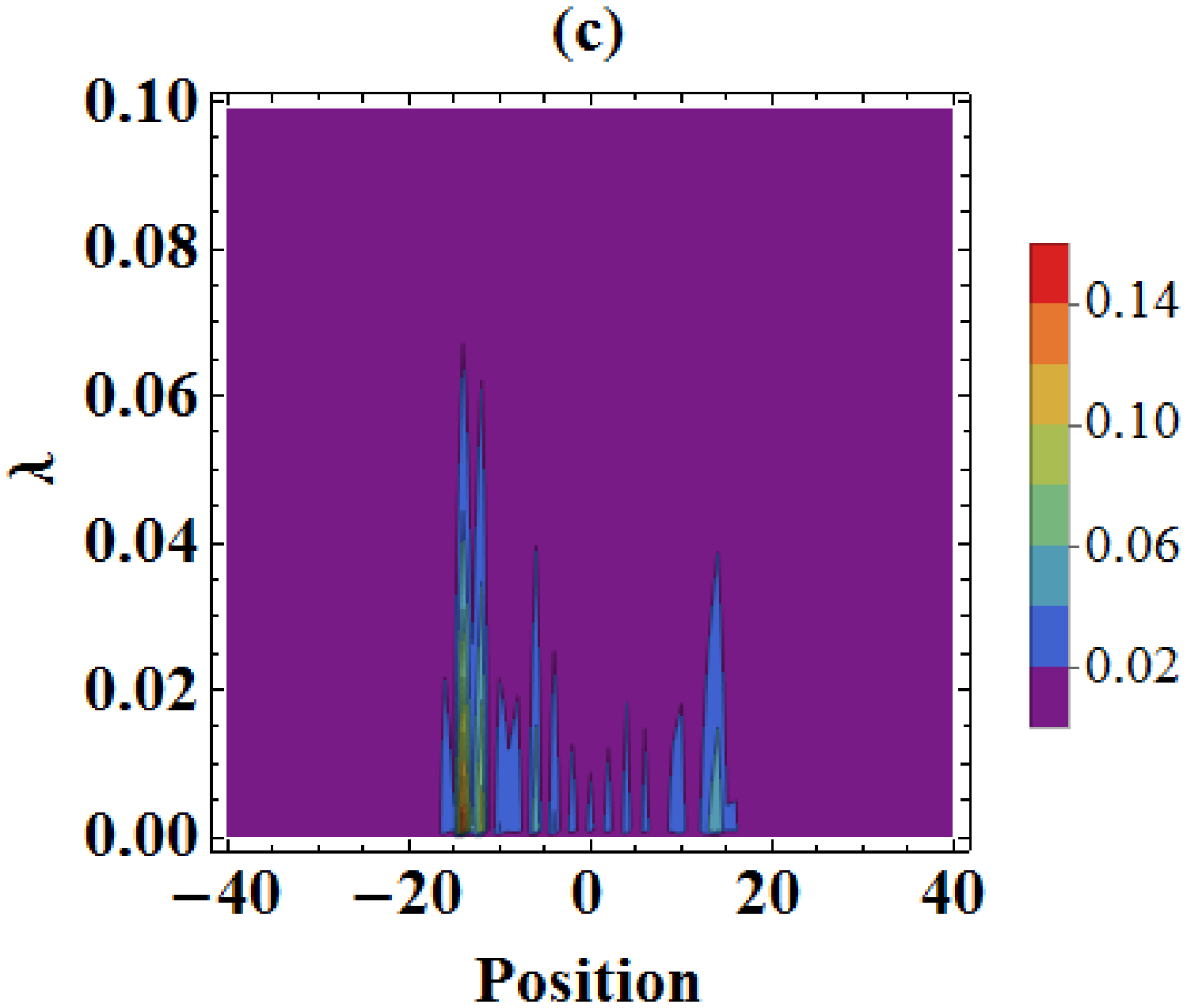}}\vspace{-1.1mm} \hspace{1.1mm}
\subfigure{\label{ab}\includegraphics[width=4.2cm]{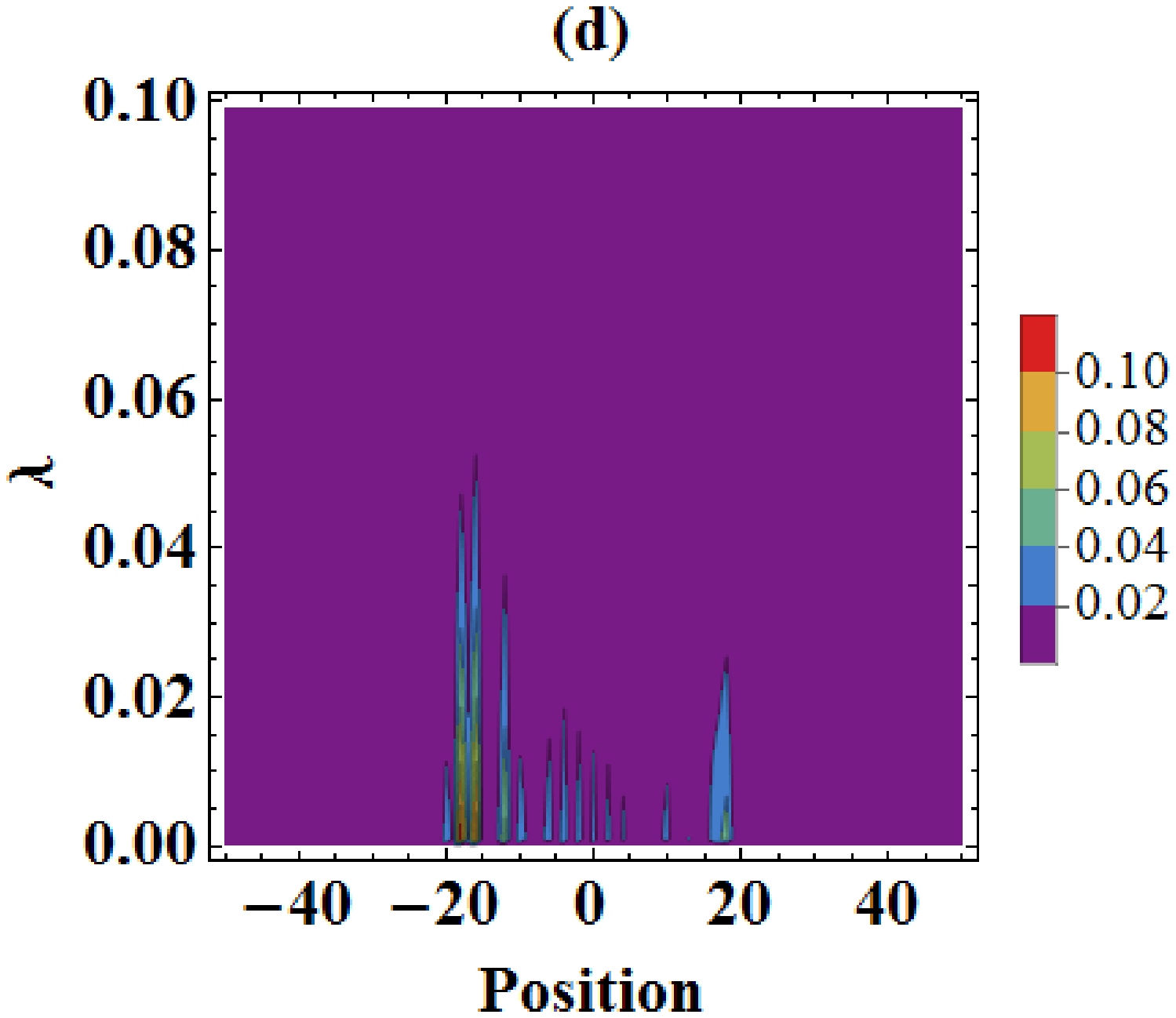}}\vspace{-1.1mm} \hspace{1.1mm}
     \end{center}
\caption{Probability distribution of a quantum walk as function of position and 
$\lambda$ obtained after  (a) 20 steps, (b) 30 steps, (b) 40 steps and (b) 50 steps.
The  coin operator in Eq.(\ref{cosH}) with parameters given in 
Eq.(\ref{sub}) is employed to generate the distribution. $\tau$ is set at 1,
with  $\hbar$=1 and  $\Omega_0$=1.
}
\label{qwalkc}
\end{figure}

In  Fig.~\ref{qwalkt}a,b, we note the spreading out of the probability distribution
with decrease in $\tau$,  and reduction in probability peaks with introduction
of a small dissipation  $\lambda$ =0.15. The reduction is most pronounced 
at larger $\tau \approx 1$. Fig.~\ref{qwalkc} highlights the  nontrivial
link between $\lambda$ and total number of steps taken by the quantum walker. 
The damping process quantified by $\lambda$ due to the leakage at each site, acts  more effectively when the
 quantum walk spans a larger number of lattice sites.

It would be worthwhile to discuss the  timescales 
involved during exciton propagation in multi-dimensional systems. Simple estimates obtained using the
random walk model by Scheblykin \cite{Sche},
show that the one-dimensional walker is required to visit 
 10$^{11}$ sites  to be quenched by a specific site in a lattice system of 10$^6$ sites.
Assuming the typical exciton lifetime of 100 ps, this implies a waiting
time,  $\tau \approx$ 10$^{-21}$s at each molecular site.
Considering the distance connecting two  molecular site to be  1 nm,   we would obtain
an exciton propagation speed of about 10$^{12}$ m s$^{-1}$ \cite{Sche},  which is higher  than the speed of
light. To evaluate the lower limit in the exciton speed,
we reduce the number of sites visited to 10$^{6}$, from which we 
 obtain typical time steps of  the order of $\tau \approx$ 10$^{-16}$s,
and an exciton speed of propagation of about 10$^{7}$ m s$^{-1}$.
These high speeds of propagation (10$^{7}$ to 10$^{12}$) m s$^{-1}$ are best described by the delocalized exciton
probing   the  crystal volume as a massively entangled system with
simultaneous feedback mechanisms. 

\subsection{Non-Hermitian coin flip operation via the exciton spin state}\label{flip}

The implementation of the coin operation via the exciton spin dynamics 
needs some explanation. One approach that can be employed to  operate
the coin flip  operation involves exploiting the spin property of exciton.
In molecular systems, the total exciton spin arising from
a combination of the electron and hole spin, takes the value of 
$S=$0 (singlet state) or $S=$1 (triplet state)\cite{thilspin}. 
The singlet state may thus be taken to represent the $|0_c\rangle$ state, while the
triplet state may correspond to the $|1_c\rangle$ state
\be
|0_c\rangle  \equiv
| {\bf{k}_e}, {\bf{k}_h}; S=0 \rangle_L = \frac{1}{\sqrt 2} \left [a_{\bf{k}_e}^\dagger(\frac{1}{2})\;
c_{\bf{k}_{h}}^\dagger(-\frac{1}{2})-
a_{\bf{k}_e}^\dagger(-\frac{1}{2})\;
c_{\bf{k}_{h}}^\dagger(\frac{1}{2}) \right ]|0_v \rangle
 \ee 
 \bea
\label{eq:opera2}
|1_c\rangle  \equiv
| {\bf{k}_e}, {\bf{k}_h}; S = 1 \rangle_L
= \frac{1}{\sqrt 3} \left ( a_{\bf{k}_e}^\dagger(\frac{1}{2})\;
c_{\bf{k}_{h}}^\dagger(\frac{1}{2}) + 
a_{\bf{k}_e}^\dagger(-\frac{1}{2})\;
c_{\bf{k}_{h}}^\dagger(-\frac{1}{2}) \right )\;
+ \nonumber \\
\frac{1}{\sqrt 6} \left (a_{\bf{k}_e}^\dagger(-\frac{1}{2})\;
c_{\bf{k}_{h}}^\dagger(\frac{1}{2})+
a_{\bf{k}_e}^\dagger(\frac{1}{2})\;
c_{\bf{k}_{h}}^\dagger(-\frac{1}{2}) \right )  |0_v \rangle
 \eea
where $|0_v \rangle$ denotes the state vector of the vacuum space
in which the electron and hole states remain in their respective ground states.
The hole (electron) creation operator is denoted by  $c_{\bf{k}_{h}}^\dagger$
 ($a_{\bf{k}_e}^\dagger(\sigma)$) with  wavevector $\bf{k}_h$ ($\bf{k}_e$)
and spin, $\sigma$. It can be seen that the  $S = 0$ spin state of the exciton 
arises from the coupling of an electron with a 
hole  of opposite spin, while the $S = 1$ spin state of the exciton 
results from the combination 
of three symmetrical spin functions of the electron and 
hole spin states with spin components, $S_{z} = +1, 0, -1$.
The $S = 1$ exciton can also be formed due to  
components,   $S_{z} = +1$ and $-1$ in the  coupling between 
electron and heavy-hole spin states, but we omit this possibility for brevity
of discussion on exciton spin states.

For practical implementations of the coin-flip operation, the differences between the singlet and
triplet excitons have to be taken into consideration. 
The energy of the triplet
state  is less than the singlet state  by 0.5 eV, and have comparatively
long diffusion lengths ($\approx 5- 10 \times 10^{-6}$m) in organic systems.
Moreover energy transfer via Coulomb coupling does not take place in triplet states
as the triplet excited state-singlet ground state
transition is spin-forbidden. Triplets thus
transfer their energy through exchange coupling instead,
accordingly the singlet exciton moves faster and  decays
quickly as well, as it is not subjected to the 
 spin-forbidden  decay rule of the triplet state. 
These differences would introduce ``unfairness" during the coin flip operation, 
which  can be minimized if the flip operation is performed within very short
time periods lasting 1 fs or less. 

The  robustness of the coin flip operation ${\hat C'}$
(Eq.(\ref{cosH})) may require further investigation of the
Elliot-Yafet (EY) momentum scattering,   Elliot-Yafet (EY),
Bir-Aroniv-Pikus (BAP) and Dyakonov-Perel (DP) mechanisms, all of which  
are known to cause spin relaxation of  charge carriers \cite{tit}.
Exciton scattering involving spin-flip processes 
due to acoustic phonons  may also
result in spin relaxation processes. A combination of these processes may
degrade the  coin flip operation ${\hat C'}$, the decrease in coin flip functionality
analyzed by information theoretic measures such as fidelity.

\subsection{Von Neumann entropy in the vicinity of the exceptional point}\label{von}

\begin{figure}[htp]
  \begin{center}
\subfigure{\label{aa}\includegraphics[width=7.2cm]{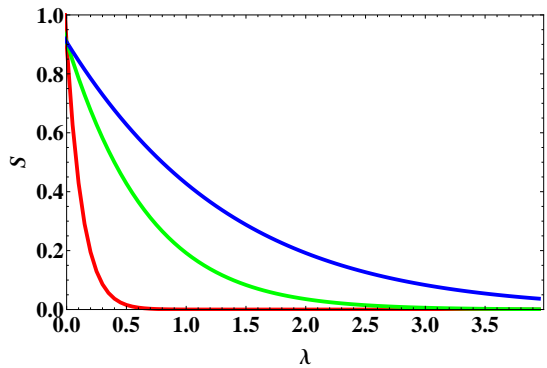}}\vspace{-1.1mm} \hspace{1.1mm}
     \end{center}
\caption{Average von Neumann entropy of a quantum walk as a function of 
$\lambda$ obtained for   $\tau=\frac{1}{5}$ (red line),  $\tau=\frac{1}{25}$ (green line)
and $\tau=\frac{1}{50}$ (blue line). The number of steps taken is fixed at 50 in all three cases.}
\label{qwalkd}
\end{figure}

The influence of the  time taken to make one quantum walk
$\tau$ on the  dynamics of the quantum walk can be further explored  via the
widely used von Neumann entropy measure \cite{niel}. We modify the expression for
the average entropy
\be
\label{ent}
S = \frac{1}{N}
\sum_{n=1}^{N}  -{\rm Tr}_n \rho_n \ln \rho_n
\ee
by setting $N$ as the total number of steps taken by the quantum walker.
$\rho_n$ is determined using the probability of occupation at site $n$. 
A large entropy, $S$  implies substantial
entanglement between various sites, while small values of 
$S$ indicates the presence of a  localized state. The entropy 
 measure is therefore useful in characterizing  quantum phase
transitions at the localization-delocalization boundary.
The von Neumann entropy measure can reveal variations in 
 entanglement properties as the exciton propagates in a dissipative
environment as shown in  Fig.~\ref{qwalkd}.
The figure shows the gradual decrease in entropy associated with increased localization
at large $\tau$, the time taken to make one quantum walk. This is expected
 as the longer time spent at a site results in greater dissipation of
energy  via a combination of processes that determine the coherence of the exciton state.
Processes which govern the excitonic coherence may include
exciton-phonon interactions, decoherence due to lattice vibrations at increased
temperatures, spin relaxation mechanisms and Pauli interactions \cite{pauli}.
Fig.~\ref{qwalkd}  shows that  $S$ vanishes at the exceptional point ($\lambda$ =4) which reflects 
the  increased localization effects  in the vicinity of
  topological defects. These results indicate that a small $\tau$ correspond to 
more efficient exciton propagation in photosynthetic
systems and show that  dissipative sinks 
favor the formation of topological defects.

\section{Exciton dynamics and quantum  tomography}\label{meaQW}
Quantities representing entanglement  are  determined using 
quantum states obtained via  quantum tomographic methods \cite{niel,tomog}.
The term ``tomography" is associated with techniques of estimating quantum states
to a high degree of accuracy via  reconstruction of density matrices.
In quantum state tomography (QST),  density matrices are estimated using  measurement techniques
that provides information about the quantum state which may be in a pure or mixed state.
By repeated measurements on quantum systems,  the derived frequency counts can be used to 
compute probabilities
that assist in building  the density matrix of a quantum state.
In quantum process tomography (QPT),  a priori quantum states 
are employed  to investigate quantum processes.
Several works  \cite{otfried,otfried2,audenaert}
have focussed on quantum measurements which can be used to retrieve some or 
all of the   information about a quantum state that is sufficient 
to quantify entanglement under specific conditions. For the 
quantum walker in which  coin and position states are entangled,
we evaluated entanglement through the Von-Neumann
entropy in Fig.~\ref{qwalkd}. It was  shown that  small values of the walk time, $\tau$,
and  damping factor, $\lambda$ can contribute to 
  conditions under which entanglement is maximally enhanced. It remains to be
seen whether these parameters play important roles during quantum tomography.

The crosspeaks in 2D electronic spectra  are proportional to the dipole
strength cross-correlation function between   exciton states \cite{cho}. This feature provides
some advantage in tracking  excitation propagation 
 on  a femtosecond timescale \cite{caram}. Recently 
an exciton relaxation time of about  50 fs was observed  as  an
off-diagonal signature in the low-energy cross-peak region of
2D coherent  spectral results \cite{milota} of molecular aggregate systems.
Interestingly  inter-band coherence spectroscopic signatures appeared at the 
same time, and correlated exciton states were projected as  off-diagonal signals in the frequency-frequency
correlation spectra with a time resolution of  about 20 fs \cite{milota}. 
These temporal features on a femtosecond timescale in the
 sequential 2D electronic spectra highlight the   
possibility of reconstructing  time evolutions of the exciton
density matrix  in the  quantum kinetics regime. The reconstruction  procedure constitutes an important
aspect of  quantum process tomography (QPT) \cite{wein,walm,excitomo}, during which  details of 
the input/output quantum state amplitudes in the
excitonic subspace are elucidated by combining a range of  experimental results.
The final outcome is subsequently derived  by  varying  polarization and frequency components of input pulses.

We note that quantum information processing offers the distinct advantage of dealing with
 exciton basis that may span over a wide and possibly intricately connected network of molecular sites.
This occurs in large   photosynthetic membranes, which may link many
biomolecular chromophores,  and possibly thousands of entangled excitonic qubits
in the Hamiltonian subspace. To this end, we focus on  two signatures of 
exciton dynamics that may be observed via  a  quantum information theoretic
 approach to spectroscopic experiments. 
The first attribute  useful to understanding 
the  correlated dynamics in excitonic systems relates to  the  presence
of topological defects such as exceptional points. In a recent work \cite{thila1}, we showed
 the appearance of exceptional points in non-Hermitian
excitonic dimers during  excitation energy transfer, for a select range of temperatures
and damping parameters. The second attribute involves the salient feature of non-Markovianity,
whose measures are based on  deviations from the 
continuous, memoryless, completely positive semi-group feature of Markovian evolution of quantum 
systems \cite{sudar,choi}.
Non-Markovian dynamics appear to be a necessary aspect of quantum dynamics of open quantum systems
when the commonly used  Markovian  model  breaks down  in the presence of
strong system-environment coupling regime or preexisting  unfactorized initial conditions 
between the system and environment. 

The role of the exceptional point and non-Markovian dynamics can be examined in the context
of a quantum walk governed by a non-Hermitian coin flip operation. During an excitonic quantum walk, the number of qubits that exist within the total Hilbert space, ${\cal H}_w$=${\cal H}_p\otimes{\cal H}_c$ increases beyond
computational tractability of non-Markovian features, even for steps taken over five or more   bacteriochlorophyll
(BChl)a molecules sites of the  FMO complex  monomer subunit. 
Moreover the number of quantum measurements required to tract the qubit dynamics also increases steeply with
the number of qubits. We therefore confine our analysis to the
quantum walk which ends at $N <$ 6 steps. 

\begin{figure}[htp]
  \begin{center}
\subfigure{\label{aa1}\includegraphics[width=6.3cm]{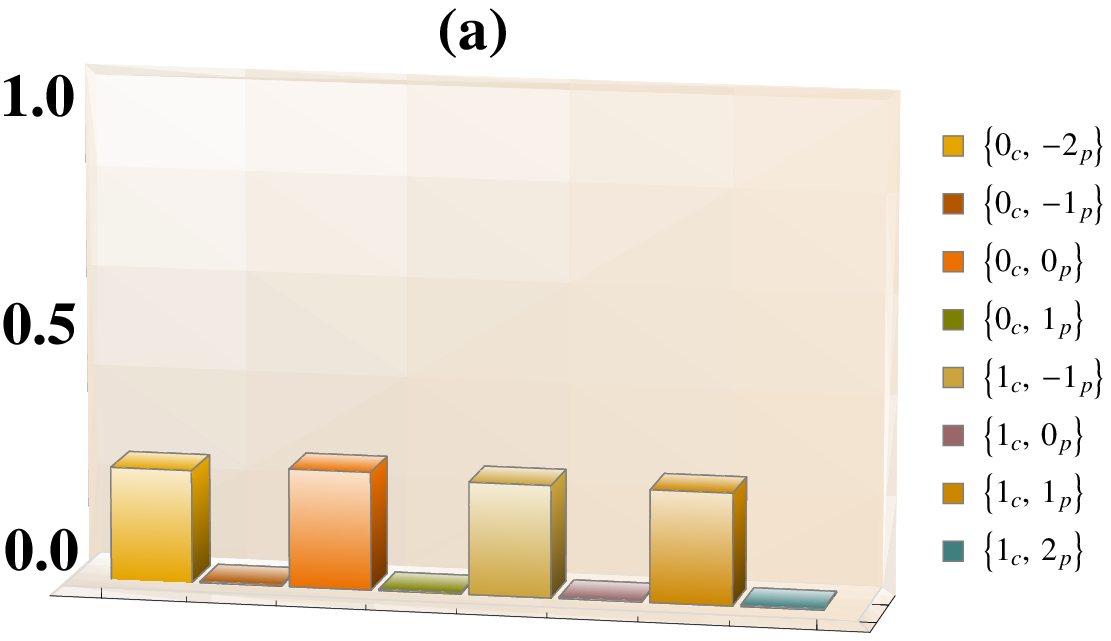}}\vspace{-1.1mm} \hspace{1.1mm}
\subfigure{\label{ab1}\includegraphics[width=6.3cm]{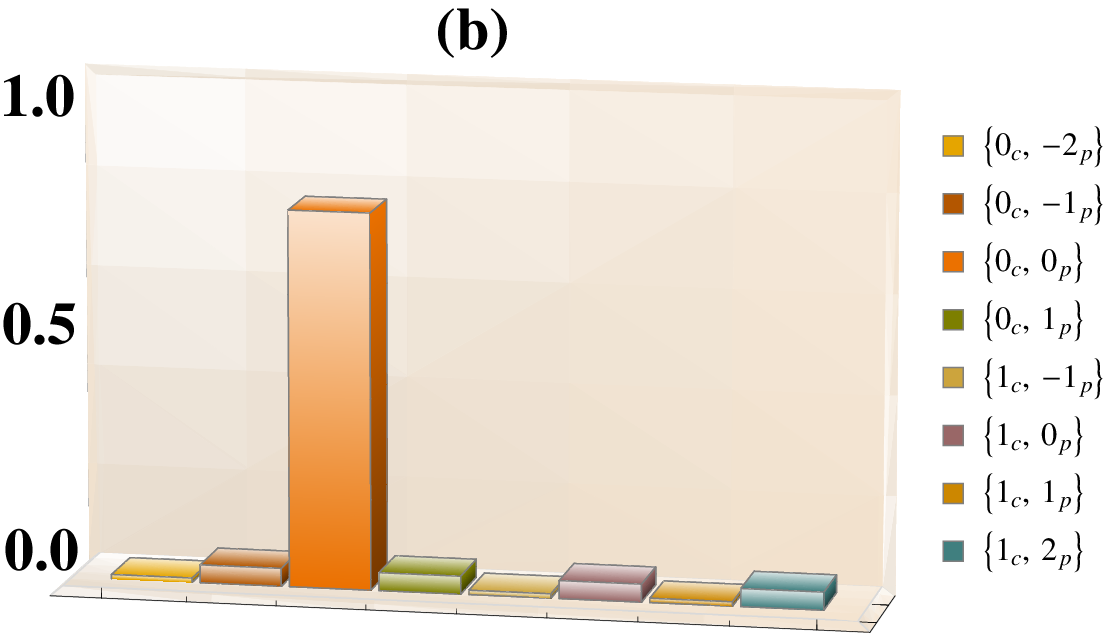}}\vspace{-1.1mm} \hspace{1.1mm}
\subfigure{\label{aa1}\includegraphics[width=7.3cm]{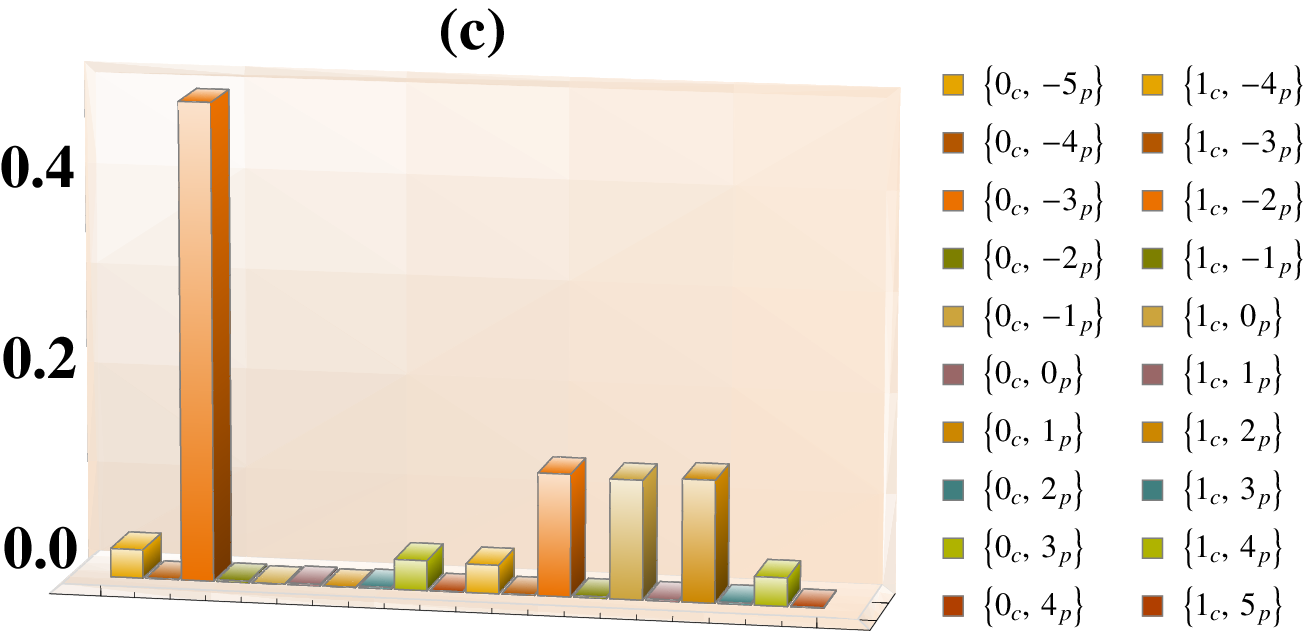}}\vspace{-1.1mm} \hspace{1.1mm}
\subfigure{\label{ab1}\includegraphics[width=7.3cm]{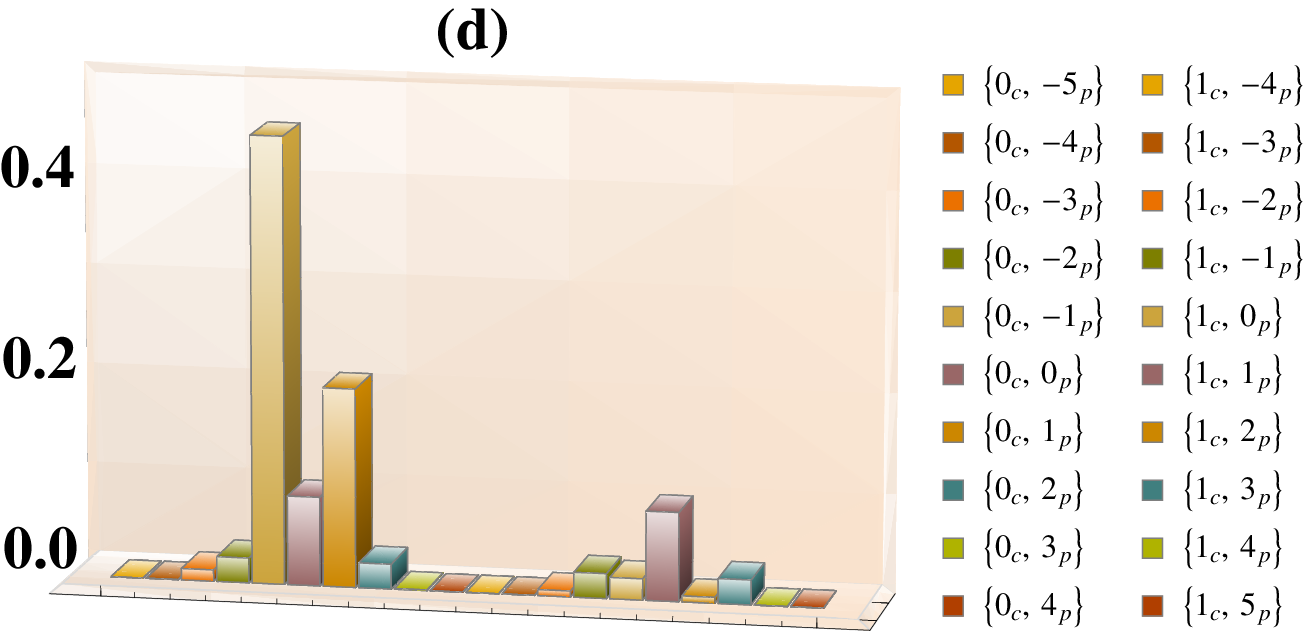}}\vspace{-1.1mm} \hspace{1.1mm}
     \end{center}
\caption{Measurement profiles of various quantum 
states as recorded by the monitoring system for $\lambda \approx$ 3.9, 
 in the vicinity of the exceptional point for (a) $\tau$=0.1, (b) $\tau$=0.5 evaluated for a quantum walk that terminates after
2 steps, and (c) $\tau$=0.1, (d) $\tau$=0.5 evaluated for a quantum walk that terminates after
5 steps.  A unit system in which $\hbar$=1,  $\Omega_0$=1 is employed. Units for time $\tau$ (time taken to make a single step)  is based on the  inverse of 
 $\Omega_0$ (at $\lambda$ = 0). $\tau$ can be interpreted as inversely proportional to the exciton diffusion speed.
}
\label{mexcep}
\end{figure}

During quantum measurements of a system $\mathcal{S}$, values of a physical quantity is transferred to a monitoring device
$\mathcal{M}$, with the mapping $\mathcal{S} \rightarrow \mathcal{M}$ representing a mathematical description of
the measurement. It is considered  that the  quantum
states of the monitoring device can be easily read, and that these quantum states are orthogonal
to one another, with fulfillment of the normalization condition. 
There are several ways to extract information from the coin-position composite system state arising
from the quantum walk. For instance, we may first perform a measurement on the 
coin using the observable of the form ${\hat O}_c = \gamma_0 |0\rangle_{c}\langle 0| + \gamma_1
|1\rangle_{c}\langle 1|$. A second measurement is  next performed on the position states  via  the operator
$ {\hat O}_p = \sum_i \alpha_i |i\rangle_{p}\langle i| $.
In Table~\ref{tab1}  we provide results of 
quantum measurements of the quantum walk after   $N=$2  steps, obtained using 
``Quantum", the package for Dirac Notation in Mathematica. Since only  the diagonal
elements are retrieved, there is insufficient information to fully characterize the entanglement
and non-classical correlation aspects of the exciton dynamics.
The diagonal terms however, will suffice to provide
useful information related to the exceptional point. 
\begin{table}[ht]
\be
\begin{array}{|c|c|c|}
\hline
 \text{Probability} & \text{Measurement} & \text{State} \\
\hline
 \frac{{\alpha_1}^4}{4 \beta^2} &\{
\begin{array}{cc}
 0_{{c}} & -2_{{p}} \\
\end{array}
\} & |0_c\rangle \otimes |-2_p\rangle  \\
\hline
 \frac{{\alpha_1}^2 {\alpha_2}^2}{4 \beta^2} & \{
\begin{array}{cc}
 0_{{c}} & -1_{{p}} \\
\end{array}
\} & -|0_c\rangle \otimes |-1_p\rangle \\
\hline \frac{1}{\beta^2}
{\left(\frac{{\alpha_1}^2}{2}+{\alpha_2}^2\right)^2}& \{
\begin{array}{cc}
 0_{{c}} & 0_{{p}} \\
\end{array}
\} & |0_c\rangle \otimes |0_p\rangle  \\
\hline
 \frac{{\alpha_1}^2 {\alpha_2}^2}{4 \beta^2} & \{
\begin{array}{cc}
 0_{{c}} & 1_{{p}} \\
\end{array}
\} & |0_c\rangle \otimes |1_p\rangle  \\
\hline
 \frac{{\alpha_1}^4}{4 \beta^2} & \{
\begin{array}{cc}
 1_{{c}} & -1_{{p}} \\
\end{array}
\} & |1_c\rangle \otimes |-1_p\rangle  \\
\hline
 \frac{{\alpha_1}^2 {\alpha_2}^2}{4 \beta^2} & \{
\begin{array}{cc}
 1_{{c}} & 0_{{p}} \\
\end{array}
\} & |1_c\rangle \otimes |0_p\rangle  \\
\hline
 \frac{{\alpha_1}^4}{4 \beta^2} & \{
\begin{array}{cc}
 1_{{c}} & 1_{{p}} \\
\end{array}
\} & -|1_c\rangle \otimes |1_p\rangle  \\
\hline
 \frac{{\alpha_1}^2 {\alpha_2}^2}{4 \beta^2} & \{
\begin{array}{cc}
 1_{{c}} & 2_{{p}} \\
\end{array}
\} & -|1_c\rangle \otimes |2_p\rangle  \\
\hline
\end{array}
\ee
\caption{Outcome of quantum measurements after two steps of a quantum walk. $\beta^2$=$\left({\alpha_1}^2+{\alpha_2}^2\right)^2$
where $\alpha_1,\alpha_2$ appear as elements in the  non-Hermitian  coin flip operator ${\hat C'}$
(see Eq.(\ref{cosH})).  The degree of dissipation of the system undergoing
non-Hermitian dynamics is determined by the difference
$\Delta$ = $\alpha_1-\alpha_2$.}
\label{tab1}
\end{table}
We use as basis the exemplary  
exciton transfer process between two sites  as described by the total Hamiltonian in Eq.(\ref{pheq}),
and the occupation probabilities in Eqs.(\ref{sub}). We evaluate the measurement based 
 profiles of various quantum 
states as recorded by the monitoring system (shown in 
Fig~\ref{mexcep}a,b) in the vicinity of  the exceptional point which occurs
when the dissipative measure, $\lambda \rightarrow$ 4. The probabilities are based
on the results of the quantum measurements  given in Table~\ref{tab1},  at two different times 
needed to make a  quantum step.
The  profiles show the increased probability (due to localization) to remain at the starting point of quantum walk
for longer time steps ($\tau$=0.5) as compared to shorter time steps($\tau$=0.1).
As expected,  a  different measurement profile (Fig~\ref{mexcep}c,d)
is obtained when the outcome is recorded after five steps of the quantum walk.
These results indicate that the conditions under which 
topological defects occur could be determined by comparing differences
in site occupation probabilities as the environmental settings are varied.
Exceptional points are generally associated with a range of system parameter attributes, 
and thus a spectrum of defects may be detected  through experiments.
This approach provides an indirect technique of  probing the quantum dynamical processes
and transitions which may occur in the femtosecond timescale,  as the typical $\tau \approx$ 1 fs.

There is also  possibility of observing  phase transitions 
via  time series-type  tomographic techniques
using  ultrafast spectroscopy experiments. The actual implementation of
experimental techniques is beyond the scope of this work as
measurement procedures are non-ideal \cite{thilam} and
the density matrix of a probed system is susceptible to noise related errors.
 One may employ algorithmic procedures such as the   Maximum Entropy
estimation \cite{buzek}, or the 
Variational Quantum Tomography (VQT) \cite{maciel} which involves the 
minimization of  a linear cost function. These approaches may help in the  estimation 
of density matrices with higher accuracy,  and the dissipative conditions under which exceptional points
and phase transitions occur.

\subsection{Non-Markovianity during quantum walk}\label{nm}
We examine the complex interplay  of non-Markovian signatures that occur in the total Hilbert space,
when quantum measurements are performed during the quantum walk.
The  regeneration  of a density matrix from an unknown state
involves  measurement projections of
 density matrices which are  Hermitian, of unity trace, and which
obey the positivity conditions \cite{sudar}. The density  matrix is obtained by projecting the 
 unknown quantum state onto a set of known states derived from experimental
results.  However the interference effects between coin and walker coupled with
interaction of the coin with the environment, may  give rise to  non-Markovian effects 
 which violates positivity conditions during the quantum walk.
 
The  trace distance $D[\rho_1,\rho_2]$  \cite{bre}
is a well known metric measure of distinguishability of two quantum states $\rho_{1},\ \rho_2$
and is computed using $D[\rho_1,\rho_2]=\frac{1}{2}||\rho_1-\rho_2||$ \cite{niel}
where $||A||=Tr[\sqrt{A^\dagger A}]$  between   $\rho_{1},\ \rho_2$.
During Markovian  evolutions,  the trace-distance  does not increase with time and 
 $D[\rho_1(t),\rho_2(t)] < D[\rho_1(0),\rho_2(0)]$. The violation of this inequality
  signals  the breakdown of the complete positivity condition during  
the time-evolution  of a quantum system. It is to be noted that 
an increase of trace distance during 
a  time intervals is a sufficient  but not necessary indication of the 
occurrence of non-Markovianity. In this work, we employ the trace-distance difference
\be
\label{traceD}
D(T, \tau) = D[\rho(\tau'),\rho(\tau)] - D[\rho(T+\tau'),\rho(T+\tau)],
\ee
where $\tau' << \tau$, to identify non-Markovian dynamics during the entangled quantum walk.

For computational simplicity, we consider the case of the 
quantum walk which terminates after  $N =$ 2 steps.
By tracing  out the coin states of the walker, we obtain
a   reduced density matrix describing the quantum walk
in the position basis $(\ket{-2_{p}},\ket{-1_{p}},\ket{0_{p}},\ket{1_{p}},\ket{2_{p}})$
\bea
\nonumber
\rho_p(\alpha_1,\alpha_2)=  
\left(
\scalemath{0.8}{
\begin{array}{ccccc}
\small
 \frac{\alpha _1^4}{4} & -\frac{1}{4} \alpha _1^3 \alpha _2 & \frac{1}{2} \alpha _1^2 \left(\frac{\alpha _1^2}{2}+\alpha _2^2\right) & \frac{1}{4} \alpha _1^3 \alpha _2 & 0 \\
 -\frac{1}{4} \alpha _1^3 \alpha _2 & \frac{\alpha _1^4}{4}+\frac{1}{4} \alpha _2^2 \alpha _1^2 & \frac{1}{4} \alpha _1^3 \alpha _2-\frac{1}{2} \alpha _1 \alpha _2 \left(\frac{\alpha _1^2}{2}+\alpha _2^2\right) & -\frac{\alpha _1^4}{4}-\frac{1}{4} \alpha _2^2 \alpha _1^2 & -\frac{1}{4} \alpha _1^3 \alpha _2 \\
 \frac{1}{2} \alpha _1^2 \left(\frac{\alpha _1^2}{2}+\alpha _2^2\right) & \frac{1}{4} \alpha _1^3 \alpha _2-\frac{1}{2} \alpha _1 \alpha _2 \left(\frac{\alpha _1^2}{2}+\alpha _2^2\right) & \frac{1}{4} \alpha _1^2 \alpha _2^2+\left(\frac{\alpha _1^2}{2}+\alpha _2^2\right){}^2 & \frac{1}{2} \alpha _1 \alpha _2 \left(\frac{\alpha _1^2}{2}+\alpha _2^2\right)-\frac{1}{4} \alpha _1^3 \alpha _2 & -\frac{1}{4} \alpha _1^2 \alpha _2^2 \\
 \frac{1}{4} \alpha _1^3 \alpha _2 & -\frac{\alpha _1^4}{4}-\frac{1}{4} \alpha _2^2 \alpha _1^2 & \frac{1}{2} \alpha _1 \alpha _2 \left(\frac{\alpha _1^2}{2}+\alpha _2^2\right)-\frac{1}{4} \alpha _1^3 \alpha _2 & \frac{\alpha _1^4}{4}+\frac{1}{4} \alpha _2^2 \alpha _1^2 & \frac{1}{4} \alpha _1^3 \alpha _2 \\
 0 & -\frac{1}{4} \alpha _1^3 \alpha _2 & -\frac{1}{4} \alpha _1^2 \alpha _2^2 & \frac{1}{4} \alpha _1^3 \alpha _2 & \frac{1}{4} \alpha _1^2 \alpha _2^2 
\end{array}}
\right)
\\
\label{matp}
\eea
where $\alpha_1,\alpha_2$ appear as elements in the  non-Hermitian  coin flip operator ${\hat C'}$ in
Eq.(\ref{cosH}). We compute the trace-distance difference (Eq.(\ref{traceD})) for a quantum walk with
a coin flip operator ${\hat C'}$ based on the exciton  occupation probabilities in Eqs.(\ref{sub})
and the reduced density matrix in Eq.(\ref{matp}).

\begin{figure}[htp]
  \begin{center}
\subfigure{\label{aa}\includegraphics[width=5.3cm]{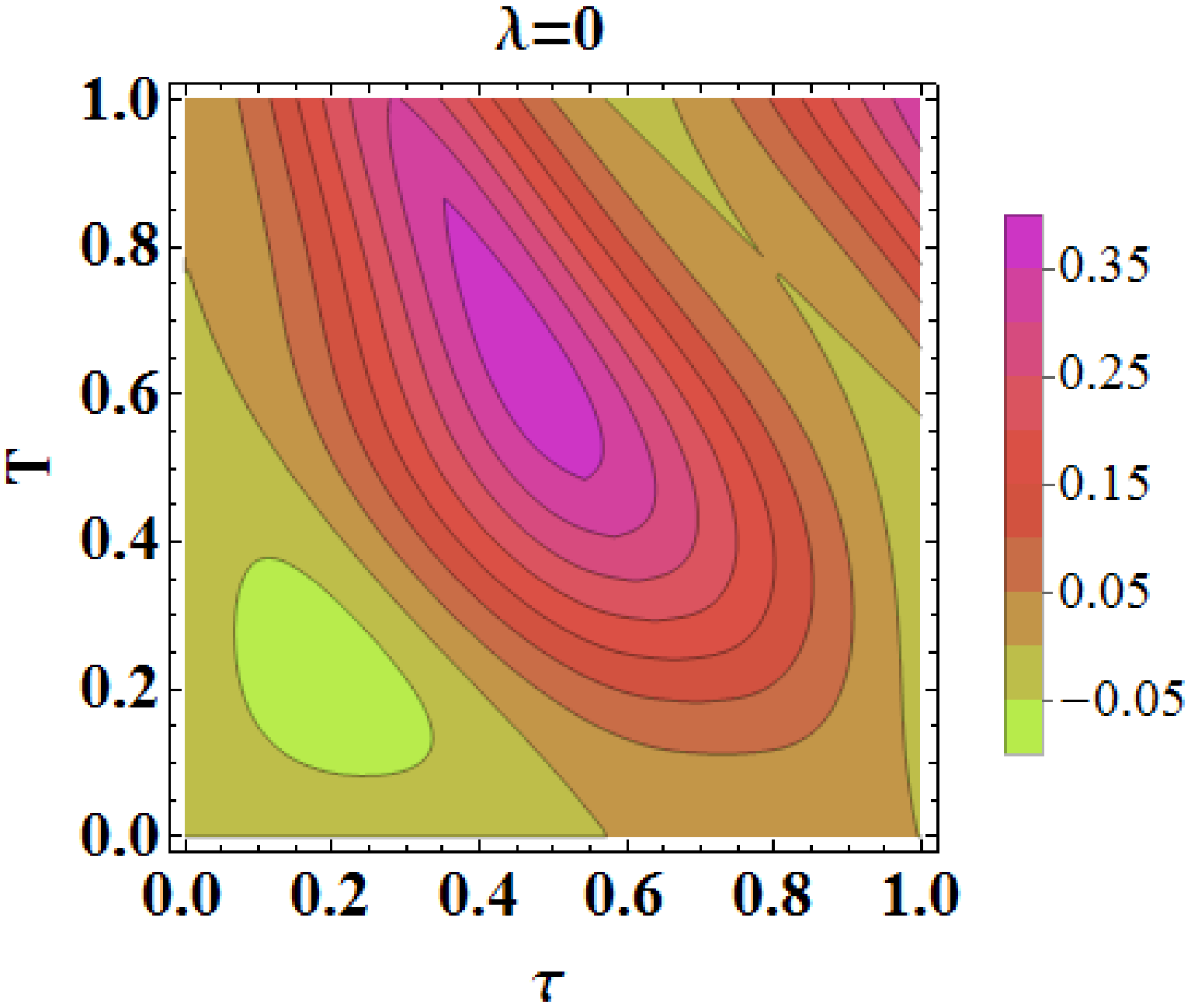}}\vspace{-1.1mm} \hspace{1.1mm}
\subfigure{\label{ab}\includegraphics[width=5cm]{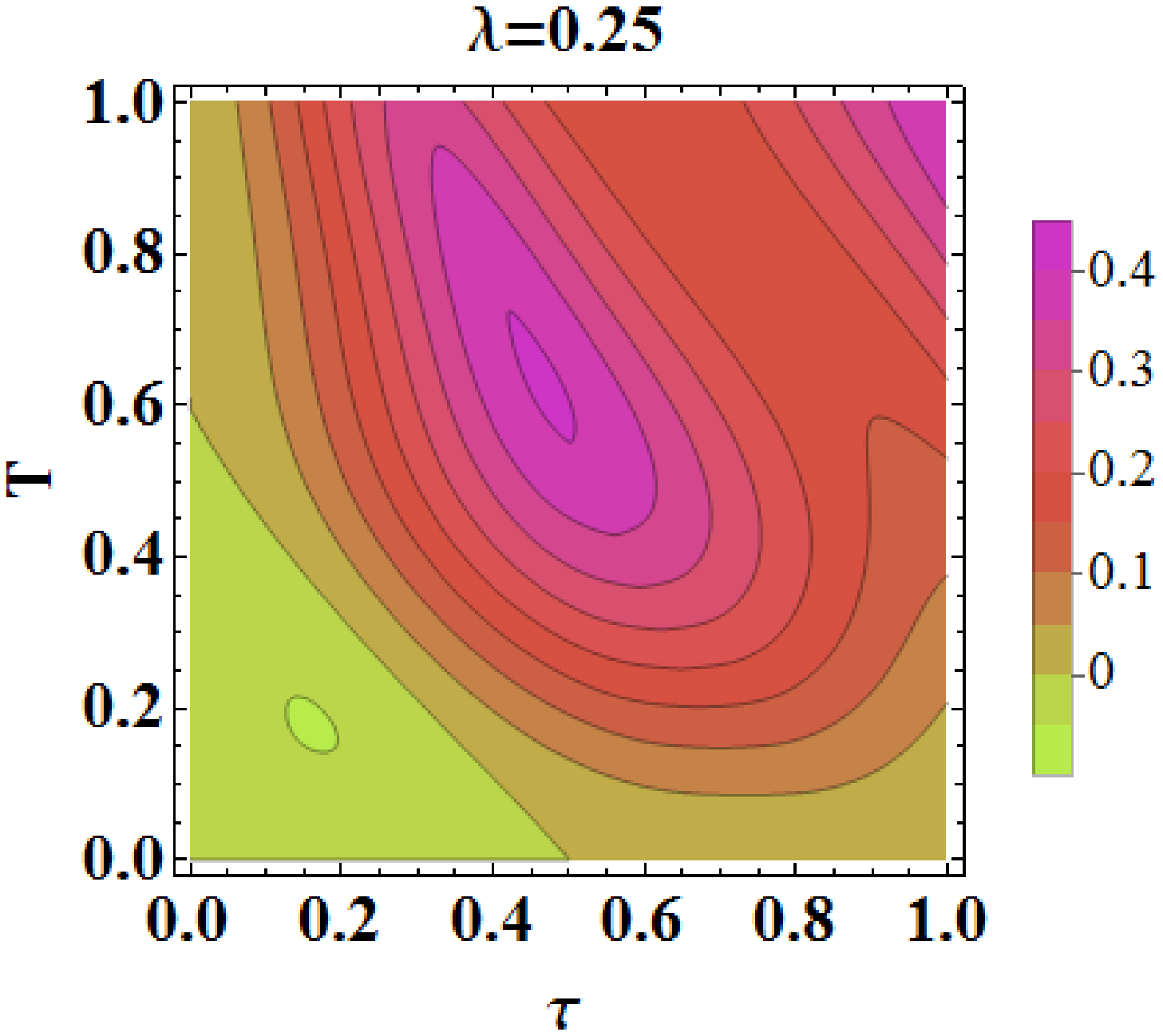}}\vspace{-1.1mm} \hspace{1.1mm}
     \end{center}
\caption{
 Trace-distance difference function,  
$D(T,\tau)$ ( Eq.(\ref{traceD}))  as a function of 
the time of one  quantum step, $\tau$ and the lapse time, $T$.
$D(T,\tau)$ is based on the reduced density matrix describing the quantum walk
in the position basis (see Eq.(\ref{matp})). The shading in the bar-legend
identifies  regions undergoing
non-Markovian dynamics (negative values of $D(T,\tau)$).
Values of  $\lambda$ are indicated at the top of each figure and
 and $\tau'$=0.00001.
 The unit system  is based on $\hbar=V=$1, with time $t$  obtained as inverse of 
 $\Omega_0$ (at $\lambda$ = 0).}
\label{qwNM}
\end{figure}

Fig.~\ref{qwNM} illustrates the trace-distance difference,  
$D(T,\tau)$ (Eq.(\ref{traceD}))  as a function of 
the time for one  quantum step, $\tau$ and the lapse time, $T$ for 
the reduced density matrix associated with a quantum walk that terminates after two steps
(see Eq.(\ref{matp})). The figure show  regions undergoing
non-Markovian dynamics (negative values of $D(T,\tau)$), at small values of $\tau, T \approx$ 0.2,
which is optimized at $\lambda$=0. 
The non-Markovian dynamics therefore dominates at the
initial period of quantum evolution,
at times of the order of the environmental bath memory time. 
The sensitivity of the qubit dynamics to the  dissipation measure $\lambda$
is highlighted by Fig.~\ref{qwNM}b. An increase in  $\lambda$  suppresses non-Markovian dynamics, 
we therefore expect  the  qubit dynamics (in position basis) to become
Markovian beyond a critical dissipation $\lambda_c$. This is 
 consistent with  diminished population exchanges between the quantum walker  and the 
dissipative environment at each lattice sites, as also seen in the decrease
of the von Neumann entropy measure  in  Fig.~\ref{qwalkd}. 
While  the results obtained in Fig.~\ref{qwNM} is
specific to the $5 \times 5$ reduced density matrix in Eq.(\ref{matp}),
similar features of information back-flow  are expected for quantum walks
terminating after higher number of steps.

Variations in  non-Markovian dynamics may appear
with further expansion of the quantum walk to two or three dimensions,
and the inclusion of greater degree of topological connectivity at each visited site.
These features are more representative of the realistic 
molecular structures in multi-chromophoric macromolecule systems.
The analysis of non-Markovian dynamics in large networks of molecular systems
would  involve  greater computational times.
To this end, it would be worthwhile to examine the possibility of tracking non-Markovian
type processes through resurgence of signatures of population dynamics in the initial quantum 
kinetic time regime via quantum tomography schemes. 
Exciton coherence structures 
are  known to be associated with quantum beat oscillations in 
the initial time period of about 100 fs \cite{milota}.
In contrast streaked signals  become superposed on the coherence frequency
patterns when the exciton relaxes without re-populating.  
Improvements in  current quantum process tomography techniques will enable
the detection of non-Markovian dynamics, and provide  better understanding of
exciton dissipative  dynamics  that underpins the high efficiencies of
energy transfer processes  in photosynthetic systems.

\section{Conclusion}\label{con}
In this paper, we examine in situ, without resorting to perturbative  or other
approximation schemes, the coherent
propagation of the one-dimensional Frenkel exciton (correlated electron-hole pair system) 
based a model of a  quantum walker in multi-dimensional Hilbert space. 
The walk is   governed by a  non-Hermitian coin flip operation
which is coupled to a generalized shift operation. 
The dissipative coin flip operation is associated with the 
loss of amplitude at newly located site, representative
of processes which occur when an exciton  is
transferred along dimer sites in  photosynthetic protein complexes.
These processes may include localization effects arising from
 diagonal disorder and decoherence effects due to thermal vibrations
at increasing temperatures. The results obtained in this study show the uniqueness of the non-Hermitian
quantum walk approach in analying exciton propagation in biomolecular systems.

We further examine the  possibilities of extracting 
topological defects such as exceptional points and non-Markovian
features utilizing  quantum  process tomography techniques.
The techniques may involve specific
 measurements of the excitonic subspace needed to
reconstruct the exciton state density matrix spanning several molecular sites,
via a series of ultrafast spectroscopic sequences,
 with simultaneous variation of
 environmental parameters. 
Entanglement and quantum correlation measures estimated
from the time-evolution characteristics of the density matrix 
will enable more accurate evaluation of the   quenching rate and  
 the exciton diffusion length in solids.

The ability to deduce topological structures, phase transitions and 
non-Markovian features, with further improvements
in optical retrieval techniques,  will certainly enhance the
 information content of  spectroscopy results.
It is expected that investigations 
focussed on  determining the links between the quantum coherence model of this work,
and  protein-induced  fluctuations and dissipation processes 
  would provide further insight to the energy transfer dynamics in correlated biomolecular
systems. The incorporation of statistical principles based on 
the linear response theory of fluctuations in the electronic energies of pigments 
which are embedded in a protein cage environment, 
and time-dependent reorganization or solvation of proteins
within a quantum walk approach,  will contribute to further 
understanding of the dynamics of  intricately connected complex 
molecular systems in the   nanoscale limit.

\section{Acknowledgments}

The author gratefully acknowledges the  support of  the Julian Schwinger Foundation Grant,
JSF-12-06-0000.  Some calculations for this work were performed using Quantum,  
A Mathematica package for Quantum calculations in Dirac bra-ket
notation. The author would like to thank Jose Luis Gomez-Munoz for 
useful correspondences regarding  this package, and Y. Shikano
for helpful comments with regards to  Refs.\cite{shi,shi2,attal}.


\end{document}